**Self-Shielded Multigroup Microscopic Cross Section Generation Using Targeted Artificial Neural Networks**

Joshua Nichols[a], Ben Whewell[b], Nathan Gibson[b], Andrew Osborne [a]*

[a]Colorado School of Mines, 1012 14th St., Golden, CO

[b]Los Alamos National Laboratory, Los Alamos, NM

*Email: osbornea@mines.edu

# Self-Shielded Multigroup Microscopic Cross Section Generation Using Targeted Artificial Neural Networks

Multigroup neutron transport models are used widely in simulation codes due to their memory and compute efficiency compared to continuous energy methods. However, multigroup codes rely on nuclear data prepared using approximate methods that require significant manual effort and expertise to account for self-shielding effects. In addition, multigroup nuclear data libraries prepared in this way can be limited in their applicability for modeling diverse reactors and operating conditions. Here we show that trained artificial neural networks can predict self-shielded microscopic cross sections accurately in pincell simulations containing uranium dioxide fuel with a fixed geometry and temperature. The model's predictive accuracy remains high across fuel enrichment and burnup ranges encountered in light water reactors. Critically, the neural networks' predictions require only the atomic concentration of the fuel's constituent nuclides as inputs. The networks are trained using 8,704 samples of training and validation data generated using continuous energy OpenMC pincell simulations. The trained neural networks predict self-shielded total, fission, absorption, elastic, and total scattering cross sections for combinations of 90 nuclides in the CASMO-8 energy group structure. The neural networks' predictions of self-shielded cross sections are validated by comparison with 1,500 samples of reference cross sections unseen during training and tallied using OpenMC simulations. The predicted and ground truth cross sections in the test data set agree to within 0.119% on average. When the predicted cross sections are used in a multigroup eigenvalue run they achieve a mean absolute error in $k_{eff}$ of 200 pcm.

## I. INTRODUCTION

Multigroup models solve neutron transport equations by discretizing the energy domain into a grid consisting of two or more groups. Simulation codes implementing multigroup methods require nuclear data tabulated on the same energy grid. While the preparation of macroscopic multigroup nuclear cross sections is straightforward when using continuous energy Monte Carlo codes, these data have limited applicability. This is because macroscopic cross sections depend on a material's atomic concentrations and the neutron flux, both of which evolve over time in an operating reactor and differ significantly between reactor types. Although microscopic cross sections can be tabulated instead and used to calculate macroscopic data at runtime, their preparation requires significantly more effort. The main difficulty in preparing accurate microscopic multigroup cross sections arises from self-shielding. This phenomenon is characterized by a reduction in the effective cross sections of a nuclide when it is present at high concentrations. Self-shielding is particularly acute in the resonance region, where large, sharp neutron absorption resonances cause the flux profile to drop accordingly. Because the energy resolution of multigroup data is often too coarse to resolve the effect of these resonances on the neutron flux, approximate methods are typically used to tabulate microscopic cross sections at various levels of self-shielding.

A variety of approximations have been developed to prepare accurate microscopic multigroup cross sections. Traditional methods range from interpolation schemes, like Bondarenko and Tone's method [1, 2], to integration-based techniques like the subgroup method [3]. Since their introduction, these methods have undergone incremental refinements that have significantly improved their applicability and accuracy [4, 5]. Nuclear data processing codes such as NJOY [6] have also been available for decades to assist in the preparation of multigroup data. Despite the versatility of these methods, preparing accurate multigroup nuclear data still requires significant manual and computational effort, as well as deep domain expertise.

Machine learning methods have been applied to numerous and diverse applications in nuclear science and engineering. A random forest model was used to choose the best group structure from a list generated by a simulated annealing algorithm for 1-D benchmark simulations [7]. An artificial neural network (ANN) was used to estimate time-dependent pressure and temperature responses to transient events in system level models of a nuclear reactor [8]. A Physics Informed Neural Network (PINN) was used to predict fuel temperature, peak cladding temperature, reactivity, and power in a loss of heat sink scenario for a nuclear battery [9]. ANNs have also outperformed conventional models when predicting nuclear binding energies [10]. A multilayer perceptron was able to predict nuclear cross section data for 7 different reactions more accurately than previous conventional methods [11].

Several machine learning methods have been applied directly to the generation of self-shielded nuclear data. In HTGR applications, ANNs have accurately predicted k-eigenvalues, reaction rates, and temperature coefficients using burnup and temperature data [12]. In another approach, a set of 295 deep neural networks were integrated into Griffin to provide online microscopic multigroup cross section generation for pebble-bed reactors [13]. ANNs have also been used to enhance the ultra-fine group method by predicting self-shielded cross sections in the resonance region of the WIMS-69 structure, utilizing fuel composition, temperature, and pin pitch as inputs [14]. Recent work [15] further demonstrated that ANNs could generate a shielded 56-group cross section library for 37 nuclides in PWR simulations, using principal component analysis to reduce dimensionality in data generated via SCALE [16]. This work was recently expanded to include cross section predictions for materials outside the fuel pin and in lattice calculations [17].

In our previous investigation [18], we demonstrated that artificial neural networks could accurately predict self-shielding factors for 8-group nuclear cross sections using training data generated by neutron Monte Carlo simulations. The present work extends this approach in two ways. First, we validate the predicted cross sections by comparing results from multigroup simulations using predicted cross sections against those using reference cross sections tallied directly from continuous-energy simulations. Second, we develop three separate neural networks, including a targeted network to resolve isolated errors in plutonium's resonance region, and increase model accuracy by including 204 additional training points to better cover edge cases. The three ANNs were trained and validated on 8,704 continuous-energy OpenMC [19] pincell simulations. Each simulation featured a unique fuel composition and included combinations of 90 nuclides. The networks predict the total, fission, absorption, total scattering, and elastic scattering cross sections in the CASMO-8 energy group structure [19]. To examine model performance, predictions were compared with a test set of 1,500 samples of reference cross sections unseen during training. The predicted and ground-truth shielded cross sections in this test set agreed to within 0.119% on average. When predicted cross sections were used in OpenMC multigroup eigenvalue simulations, the results differed from the reference multigroup simulations by an average of 200 pcm in $k_{eff}$.

## II. METHODOLOGY

### II.A. Training Data Generation

#### *II.A.1. Simulation Model Parameters*

A total of 10,204 continuous-energy OpenMC simulations of a light water reactor (LWR) pincell infinite lattice were run. From these, 8,704 samples were used as training (7,204) and validation (1,500) data, and the remaining 1,500 unique samples were reserved for testing. The geometry of the pincell is shown in Figure 1, comprising UOX fuel, zirconium cladding, a helium gap, and borated water coolant. The fuel, cladding, and coolant temperatures were set to 900 K, 600 K, and 562 K, respectively. A 2-dimensional model was employed, applying reflective boundary conditions around the coolant. The simulations were run with 10 inactive and 90 active batches, using 10,000 neutrons per batch.

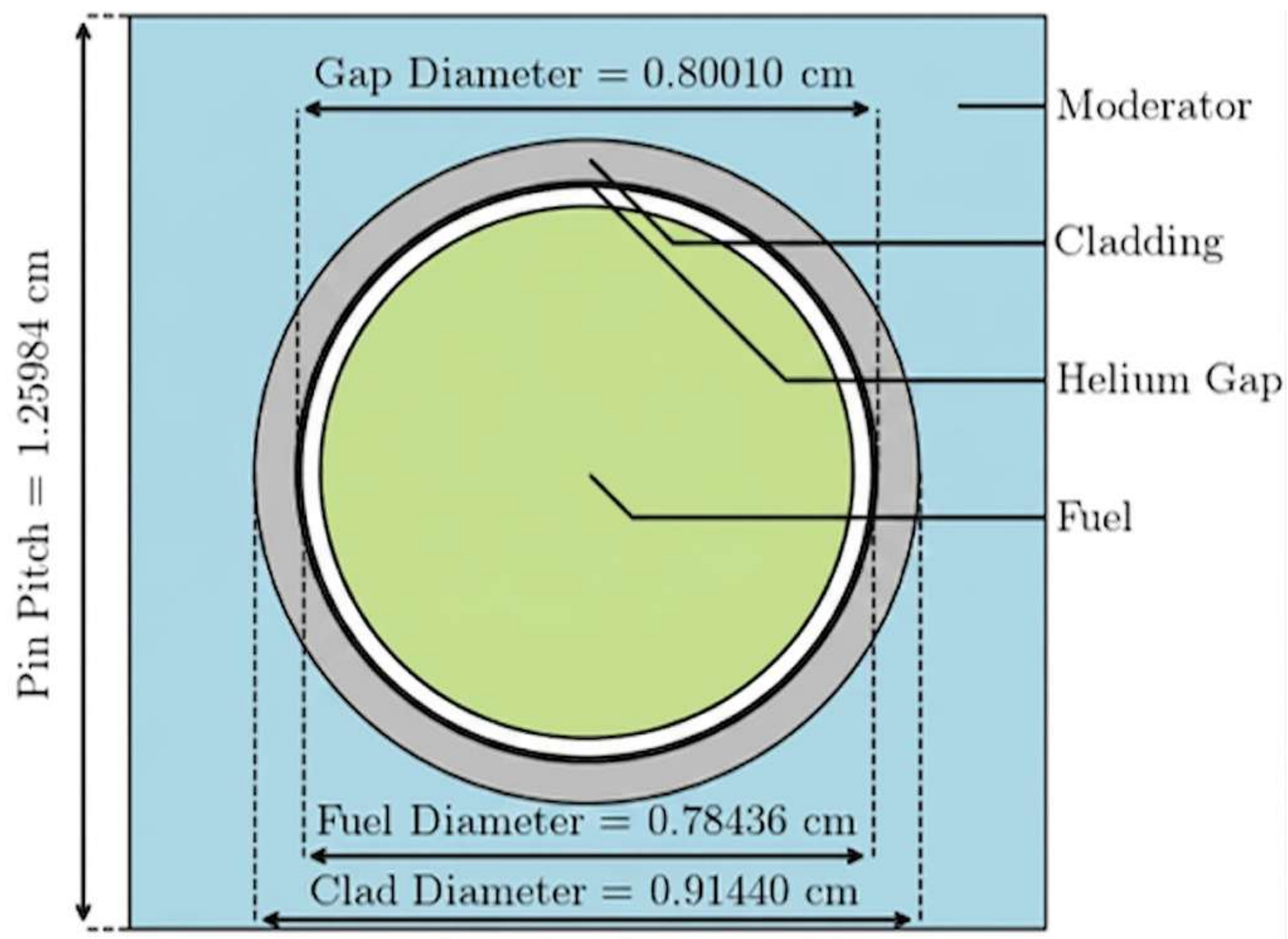

Figure 1. Pincell Geometry

*II.A.2. Fuel Composition Definition*

The isotopic composition of the fuel in each simulation was unique and selected using Latin hypercube sampling. The range of possible concentrations for each nuclide was established using previous depletion simulations of UOX fuel pins in an LWR [20]. The fuel enrichment level of those simulations ranged from 1.6% to 19.9% a/o, and they were depleted from 18 to 210 MWd/kgIHM depending on the enrichment. The upper and lower concentration limits for each nuclide were defined by finding its maximum and minimum concentrations across all depletion simulations, replacing zero concentrations in fresh fuel with trace values of $10^{-21}$ atm/b-cm. To ensure the training data covered the range of isotopic compositions expected in diverse UOX-fueled LWR designs, these maximum and minimum boundaries were expanded by ±10%.

To improve model accuracy on edge cases, 204 boundary configurations were manually included with the training data. First, each of the 90 targeted nuclides was individually set to its maximum possible concentration while the remaining 89 were held at their minimum and vice versa, resulting in 180 additional configurations. Next, the 90 nuclides were grouped into three distinct categories shown in Table 1. $^{16}$O was included in the major actinide group to better represent an oxide fuel type when concentrations of the major actinides were increased. Within these categories, concentrations were scaled to 85%, 90%, 95%, and 100% of their respective maxima while all other nuclides remained at their minima. The inverse configurations, where each category was set to its minimum while the remaining nuclides scaled to the same 85%, 90%, 95%, and 100% of their maxima, were also included resulting in an additional 24 combinations.

Table 1. Included nuclides and groupings.

| Nuclide Groups | Nuclides |
|---|---|
| **Major Actinides** | $^{234}$U, $^{235}$U, $^{236}$U, $^{237}$U, $^{238}$U, $^{238}$Pu, $^{239}$Pu, $^{240}$Pu, $^{241}$Pu, $^{242}$Pu, $^{16}$O |
| **Minor Actinides** | $^{237}$Np, $^{239}$Np, $^{241}$Am, $^{243}$Am, $^{242}$Cm, $^{244}$Cm, $^{245}$Cm, $^{246}$Cm |
| **Fission Products** | $^{142}$Nd, $^{143}$Nd, $^{144}$Nd, $^{145}$Nd, $^{146}$Nd, $^{148}$Nd, $^{150}$Nd, $^{140}$Ce, $^{141}$Ce, $^{142}$Ce, $^{144}$Ce, $^{128}$Xe, $^{131}$Xe, $^{132}$Xe, $^{134}$Xe, $^{135}$Xe, $^{136}$Xe, $^{141}$Pr, $^{143}$Pr, $^{147}$Sm, $^{148}$Sm, $^{149}$Sm, $^{150}$Sm, $^{151}$Sm, $^{152}$Sm, $^{154}$Sm, $^{139}$La, $^{133}$Cs, $^{134}$Cs, $^{135}$Cs, $^{137}$Cs, $^{95}$Mo, $^{96}$Mo, $^{97}$Mo, $^{98}$Mo, $^{100}$Mo, $^{100}$Ru, $^{101}$Ru, $^{102}$Ru, $^{104}$Ru, $^{106}$Ru, $^{99}$Tc, $^{104}$Pd, $^{105}$Pd, $^{106}$Pd, $^{107}$Pd, $^{108}$Pd, $^{134}$Ba, $^{140}$Ba, $^{90}$Sr, $^{89}$Y, $^{103}$Rh, $^{105}$Rh, $^{147}$Pm, $^{109}$Ag, $^{83}$Kr, $^{84}$Kr, $^{86}$Kr, $^{110}$Cd, $^{111}$Cd, $^{127}$I, $^{153}$Eu, $^{154}$Eu, $^{155}$Eu, $^{81}$Br, $^{154}$Gd, $^{156}$Gd, $^{158}$Gd, $^{91}$Zr, $^{93}$Zr, $^{96}$Zr |

Although the depleted fuel compositions contained 199 total nuclides, the set of predicted values was reduced to isolate the nuclides responsible for the greatest neutronic effects. These were identified based on their contribution to the one-group macroscopic total cross section and the total reaction rate of a reference material containing the maximum concentration of every nuclide. The green curve in Figure 2 shows the cumulative sum of the fuel's total reaction rate as nuclides are added in descending order of their respective total reaction rates. Here the total one group reaction rates were calculated using a single pincell infinite lattice simulation with 20,000 neutrons per batch; 100 active and 20 inactive batches. A threshold of 99.9% of the cumulative sum of each total reaction rate tallied was used to identify the 86 nuclides with the largest total reaction rate and thus the greatest impact on the neutronics. The blue curve in Figure 2 shows the cumulative sum of the fuel’s macroscopic total cross section as nuclides are added in descending order of their individual macroscopic total cross sections. In this case the same 99.9% threshold identified the 87 nuclides with the largest total macroscopic cross sections in the fuel. The total cross sections were calculated using one-group values taken from the raw ENDF data using OpenMC's data API. To rank the neutronic importance of each nuclide, the one-group cross section was calculated by collapsing the raw data assuming a flat neutron flux. The union of these sets of 87 and 86 nuclides produced 90 unique nuclides shown in Table 1, whose concentrations were chosen as input features for the neural networks.

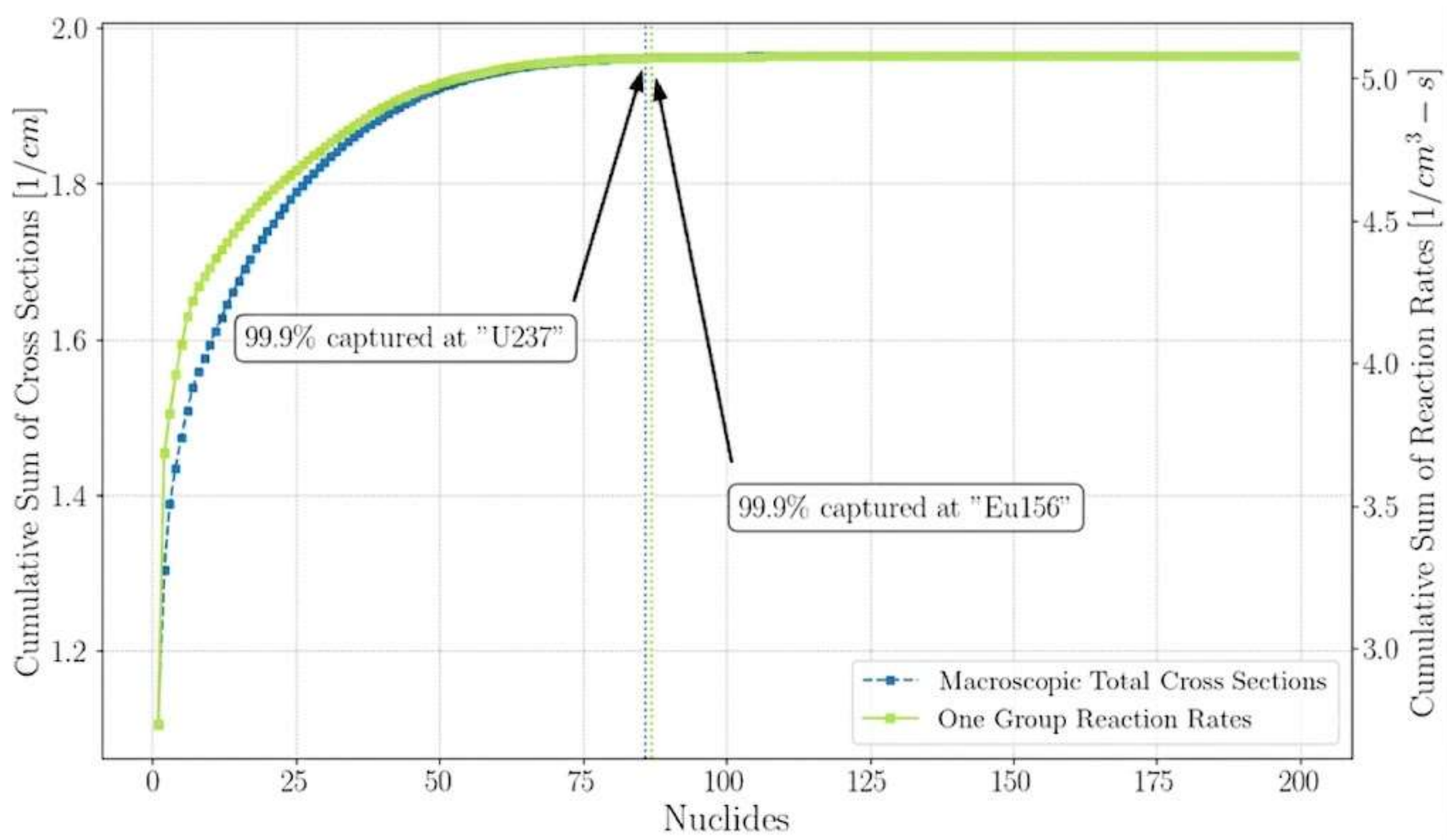

Figure 2. Cumulative sum of macroscopic total cross sections (blue) and one group reaction rates normalized per unit starting particle (green).

*II.A.3. Data Curation and Feature Engineering*

The atomic concentrations of the 90 selected nuclides served as the inputs for the neural networks. The total, absorption, fission, elastic scattering, and total scattering reaction rates, alongside their respective uncertainties, were tallied in the CASMO-8 group structure for each nuclide across all 10,204 samples. From these tallies, the effective microscopic cross sections were calculated by dividing the reaction rate by the scalar flux and atomic concentration of each nuclide, as shown in Equation 1.

$$\sigma_{\mathrm{i}} = \frac{RR_i}{\mathrm{N_i}\phi} \tag{1}$$

Where $\sigma_i$ is the effective microscopic cross section for nuclide *i*, $RR_i$ is the tallied reaction rate, $N_i$ is the nuclide concentration and $\phi$ is the flux in the fuel. The unshielded multigroup microscopic cross sections for each nuclide were calculated by collapsing cross sections obtained using OpenMC's data API to the CASMO-8 group structure using a flat neutron flux. The choice of weighting flux is inconsequential here, as these unshielded cross sections serve purely as a baseline from which the shielded data are computed as ratios. These ratios between the shielded and unshielded cross sections for 90 nuclides, 8 groups and 5 reactions then defined the high dimensional output space for the neural networks. To improve prediction accuracy, particularly because only 18 of these 90 nuclides are fissionable, the prediction workload was divided. Initially, two neural networks were trained: a Fission Neural Network (FNN) for the fission reaction in the 18 fissionable nuclides, and a Total Neural Network (TNN) for the remaining reactions. These models, however, exhibited large errors in specific epithermal groups: group 4 for $^{239}$Pu and $^{241}$Pu, and group 5 for $^{238}$Pu, $^{240}$Pu, and $^{242}$Pu. To improve predictions, a third Plutonium Neural Network (PNN) was trained exclusively on these 25 output values.

**II.B. Neural Network Architecture and Training Protocol**

The three neural networks were implemented using the PyTorch library. All three networks utilized an identical input layer of 90 neurons, corresponding to the atomic concentrations of the selected nuclides. Prior to training, these input concentrations [atm/b-cm] were normalized using scikit-learn's Min-Max scaler [21].

The Fission Neural Network (FNN) was configured with two hidden layers containing 128 and 512 neurons, respectively. Its output layer consisted of 139 neurons. This corresponds to the predicted fission cross sections for 18 fissionable nuclides across 8 energy groups, a total of 144 values, but omitting the 5 specific plutonium values delegated to the PNN.

The Total Neural Network (TNN) was tasked with predicting the total, absorption, elastic scattering, and total scattering cross sections. Like the FNN, it used two hidden layers of 128 and 512 neurons. Its output layer contained 2,860 neurons, corresponding to the 4 reactions across 90 nuclides in 8 groups, a total of 2,880 values, excluding the 20 non-fission values handled by the PNN.

Finally, the Plutonium Neural Network (PNN) was trained exclusively on the highly variable resonance outliers: group 4 for $^{239}$Pu and $^{241}$Pu, and group 5 for $^{238}$Pu, $^{240}$Pu, and $^{242}$Pu. Predicting all 5 reaction types for these specific nuclide-group combinations resulted in an output space of 25 neurons. The PNN shared the same hidden layer topology as the FNN and TNN. The remaining hyperparameters were uniform across all three models and are detailed in Table 2.

Table 2. Neural Network Parameters.

| Parameter | Description |
|---|---|
| **Activation Function** | ReLU |
| **Loss Function** | Gaussian NLL |
| **Learning Rate** | $1.0 \times 10^{-3}$ |
| **Optimizer** | Adam |
| **Batch Size** | 300 |
| **FNN Topology** | $90 \times 128 \times 512 \times 139$ |
| **TNN Topology** | $90 \times 128 \times 512 \times 2860$ |
| **PNN Topology** | $90 \times 128 \times 512 \times 25$ |

The Gaussian negative log-likelihood (NLL) loss was selected to incorporate the statistical uncertainty of the Monte Carlo reference data. Rather than predicting the variance, each reference cross section's tallied uncertainty was supplied directly, allowing the loss to down-weight targets with poor statistics. To reduce overfitting and optimize training time, early stopping was used along with a learning rate scheduler. The learning rate scheduler decreased the learning rate by a factor of 0.5 every 10 epochs if the validation loss did not achieve a new minimum. A patience counter was incremented each consecutive epoch the validation loss did not reach a new minimum. When the patience counter reached 250 the model stopped learning, but if a new minimum was reached then the counter was reset to 0. The model's weights and biases were saved at each epoch that the validation loss reached a new minimum value.

## III. RESULTS AND ANALYSIS

### III.A. Quantitative Model Performance

A summary of the models' predictive accuracy is given in Table 3. The table shows the training and test results from all models separately. The average quantities are weighted according to the number of output neurons in each network. It is important to note that the absolute error recorded represents the average absolute error of the entire training or test data sets in barns. The PNN has the largest absolute error reflecting the large magnitude of the resonance cross sections it predicts, which range in the thousands of barns.

Table 3. Neural Network Results on Training and Test Data.

| Metric | TNN | PNN | FNN | Average |
|---|---|---|---|---|
| **Training Loss** | 0.00313 | 0.00333 | 0.00367 | 0.00316 |
| **Training % Error** | 0.103% | 0.369% | 0.189% | 0.109% |
| **Training Absolute Error [b]** | 3.59 | 9.67 | 7.02 | 3.79 |
| **Test Loss** | 0.00323 | 0.00401 | 0.00387 | 0.00327 |
| **Test % Error** | 0.112% | 0.475% | 0.202% | 0.119% |
| **Test Absolute Error [b]** | 3.90 | 10.71 | 7.47 | 4.12 |

To evaluate the validity of the method, multigroup simulations were run using the predicted shielded cross sections and multigroup libraries generated by OpenMC for each of the 1,500 test cases. The results of these runs were then also compared with continuous energy baseline results. These three simulations are defined below:

1. CE Reference: The continuous-energy OpenMC simulation, serving as the absolute baseline.
2. GT-MG (Ground-Truth Multigroup): The multigroup simulation using the microscopic cross section library tallied directly from the CE Reference runs. These data include the total, absorption, and fission cross sections; the scattering matrices, multiplicity matrices, and chi vectors.
3. ANN-MG (Predicted Multigroup): The multigroup simulation utilizing the GT-MG library, but with the total, absorption, and fission cross sections replaced by the neural network predictions.

The microscopic libraries for the GT-MG and ANN-MG setups were converted to macroscopic libraries at runtime. The mean absolute $\Delta k_{eff}$ between the GT-MG and CE Reference was 225 pcm, averaged across 1,500 simulations. When using the ANN-MG setup, the mean absolute $\Delta k_{eff}$ compared to the CE Reference was 310 pcm. Lastly, the comparison between ANN-MG and GT-MG runs resulted in a mean absolute $\Delta k_{eff}$ of 200 pcm. Figure 3 illustrates the distribution of absolute $\Delta k_{eff}$ across these comparisons, where wider regions indicate a higher density of cases at that $\Delta k_{eff}$ value.

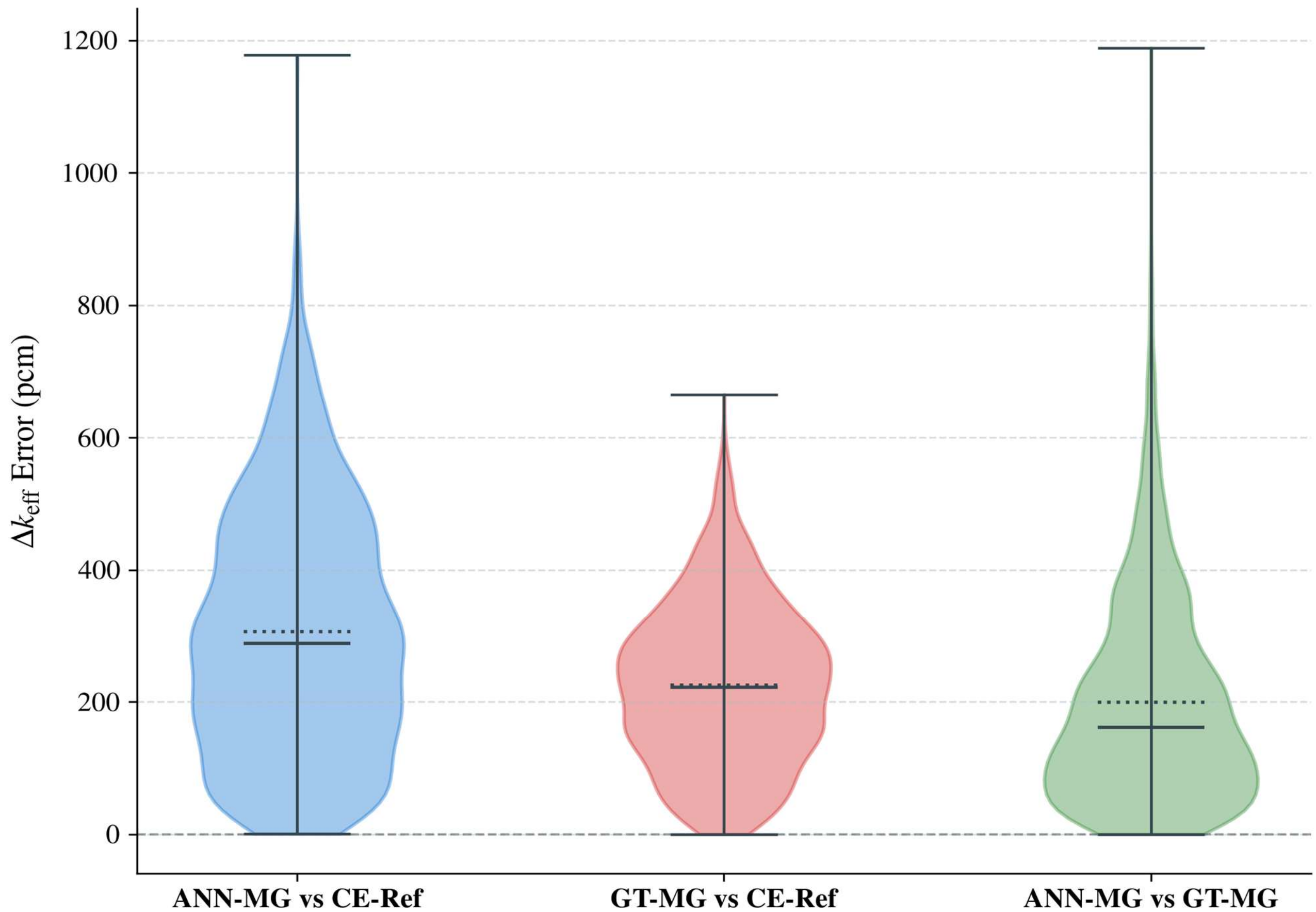

Figure 3: Distribution of absolute $\Delta k_{eff}$ (pcm) values for the three simulation comparisons. The blue plot represents the total model difference (ANN-MG vs. CE Reference), the red plot represents the baseline multigroup discretization difference (GT-MG vs. CE Reference), and the green plot isolates the difference introduced solely by the neural network predictions (ANN-MG vs. GT-MG).

Notably, the error introduced by the model alone (200 pcm) is comparable to the discretization error inherent to the multigroup approximation itself (225 pcm). Additionally, the total model error from the CE Reference (310 pcm) is smaller than the sum of the two component errors (425 pcm).

### III.B. Model Limitations

One test case exhibited a $\Delta k_{eff}$ of 1,189 pcm between the GT-MG and ANN-MG runs. While this represents the maximum deviation among the test cases, it remains well within the bounds in related published studies, which have reported maximum $\Delta k_{eff}$ approaching 3,000 pcm [17]. Furthermore, the mean absolute error of 200 pcm is lower than the published range of 398 to 965 pcm [15, 17].

Table 4 summarizes the performance metrics for the predicted cross section values whose relative errors exceed 10%, ranked by maximum relative error. These values represent 0.0021% of all model predictions, with each entry corresponding to a single nuclide and energy group.

Table 4: Predictions by nuclide and energy group with relative errors above 10%.

| Nuclide | Energy Group | Errors Above 10% | Mean Absolute Error [b] | Maximum Absolute Error [b] | Mean Relative Error | Maximum Relative Error |
|---|---|---|---|---|---|---|
| **$^{136}$Xe** | 6 | 22 | 0.0000971 | 0.00679 | 1.698% | 119.601% |
| **$^{151}$Sm** | 5 | 66 | 14.9 | 169 | 2.455% | 15.310% |
| **$^{140}$Ce** | 6 | 5 | 0.000197 | 0.00187 | 1.610% | 14.931% |
| **$^{84}$Kr** | 6 | 1 | 0.00231 | 0.0330 | 1.009% | 14.832% |
| **$^{236}$U** | 6 | 16 | 0.268 | 2.72 | 2.121% | 13.577% |
| **$^{134}$Xe** | 6 | 3 | 0.000623 | 0.00603 | 1.496% | 11.540% |

The maximum relative error in Table 4 was 119%, occurring in the group 6 absorption cross section for $^{136}Xe$. The corresponding absolute error of 0.00679 barns reflects the small magnitude of the cross section in this group. The test case containing this predicted value yielded a $\Delta k_{eff}$ of 510 pcm between the GT-MG and ANN-MG runs, which is more than twice the average $\Delta k_{eff}$ of 200 pcm. Figure 4 shows that the predicted cross section closely matches the ground truth across other groups, with the largest deviation occurring in group 6.

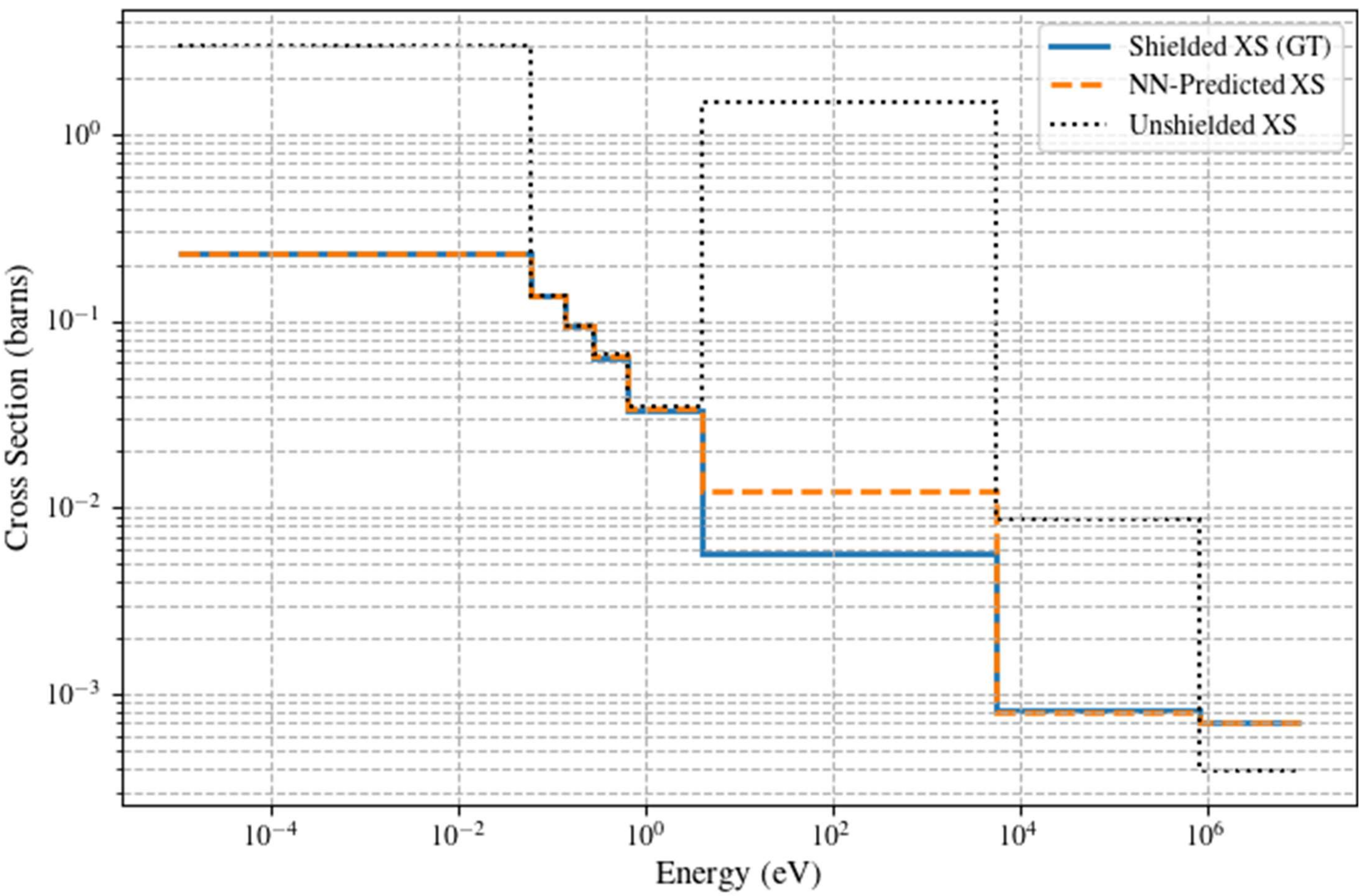

Figure 4. Eight-group absorption cross section for $^{136}Xe$, the prediction with the largest relative error. The black dotted line shows the unshielded cross section collapsed with a flat flux, the orange dashed line is the model's predicted cross section values, and the blue line is the ground truth self-shielded cross sections generated by OpenMC.

The errors in Table 4 are localized to individual groups and nuclides. Previous iterations of the TNN and FNN had similar localized errors in groups 4 and 5 of plutonium nuclides, where relative errors ranged from 10% to 20% [18]. In this work, including the Plutonium Neural Network (PNN) reduced the relative errors in plutonium nuclides to an average of 0.5% and a maximum of 7%. The same approach could be applied to correct the remaining cases in Table 4.

### III.C. Flux Spectrum Analysis

Neutron flux spectra were tallied during the multigroup and continuous-energy runs for each of the 1500 test cases. Figure 5 shows the eight-group neutron flux spectra in the fuel region for the test case with the maximum $\Delta k_{eff}$ (1,189 pcm). Even in this worst case, the spectra produced by the CE-Reference, GT-MG, and ANN-MG runs are visually indistinguishable across all eight energy groups.

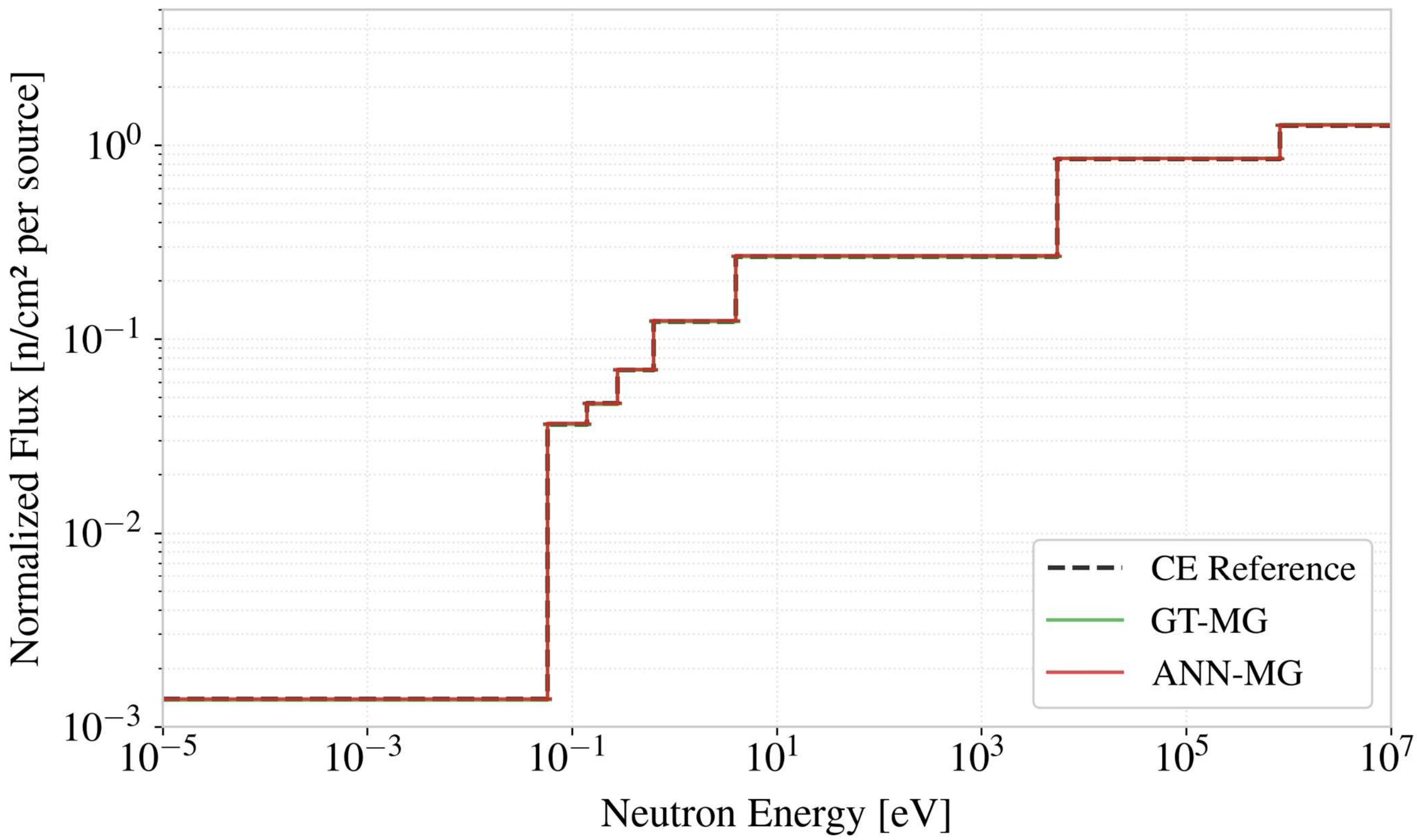

Figure 5. Eight-group neutron flux spectra normalized per unit lethargy in the fuel region of the pincell for the test case with the highest $\Delta k_{eff}$. The CE Reference (black-dotted line), GT-MG (green line), and ANN-MG (red line) flux spectra are overlaid.

The mean absolute percentage error (MAPE) of the flux spectrum between the GT-MG and ANN-MG for both the worst and best test cases were calculated. In the best test case, where the $\Delta k_{eff}$ was only 0.17 pcm, the MAPE averaged over all eight energy groups was 0.25 ± 0.08%. In the worst test case, where the $\Delta k_{eff}$ was 1,189 pcm, the MAPE was only 1.23 ± 0.07%, where reported uncertainty is the statistical uncertainty of the OpenMC flux tallies. The $\Delta k_{eff}$ values for the two test cases differ by more than three orders of magnitude, yet the flux errors differ by less than a factor of 5 and remain below 2% in both cases. This shows that even the largest observed deviation in $k_{eff}$ corresponds to only a small change in the flux spectrum, and that the flux error is the same order of magnitude as the model's cross section prediction errors.

## III.D. Model Error

To evaluate the statistical distribution of the model's accuracy across the entire test dataset, Figure 6 presents histograms of the relative errors and the pull. The relative error axis (top) is truncated between ±20% for clarity; this excludes three predictions listed in table 4, including the 119% error, though all three are included in the calculations below. The pull is the prediction residuals normalized by the standard deviation of the corresponding OpenMC ground-truth tally shown in equation 2.

$$z = \frac{x_{ANN} - x_{OpenMC}}{\sigma_{OpenMC}} \qquad (2)$$

Where z is the pull, $x_{ANN}$ is the predicted cross section value, $x_{OpenMC}$ is the ground truth cross section generated by OpenMC and $\sigma_{OpenMC}$ is the Monte Carlo standard deviation. If prediction errors are negligible, the pull distribution follows a normal distribution with a standard deviation of one. A standard deviation below one indicates the model predictions are more precise than the Monte Carlo statistical noise.

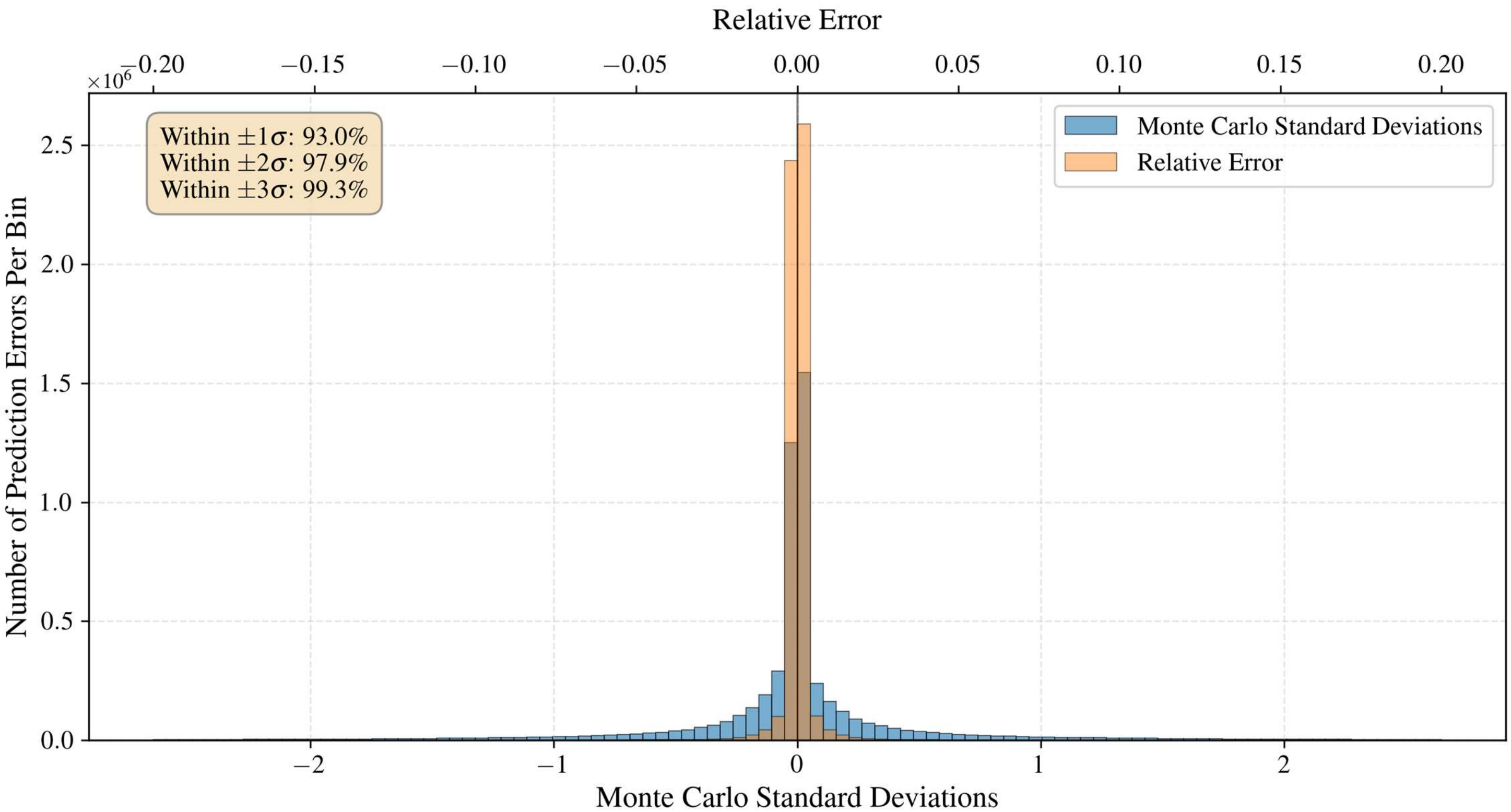

Figure 6. Histogram of the prediction errors in the test dataset. The distribution in orange represents relative errors (top axis). The distribution in blue shows the pull (bottom axis), where the prediction residuals are normalized by the Monte Carlo uncertainty.

As shown in Figure 6, 93.0%, 97.9%, and 99.3%, of the model's predictions fall within one, two, and three standard deviations of the Monte Carlo ground truth, respectively. The pull distribution has a mean of -0.01 and a standard deviation of 0.64, demonstrating that the error introduced by the model is negligible compared to the underlying statistical uncertainty of the Monte Carlo training data and has minimal bias toward under or over predictions.

**III.E. Impact of Training Data Size**

Machine learning models are susceptible to overfitting, where the training loss is significantly lower than the validation or test loss [22]. While overfitting can arise from various factors, one common cause is a large number of trainable parameters (weights and biases) relative to the size of the training dataset. To verify that the network wasn't overtraining, the training, validation and test losses were calculated. Figure 7 shows how these losses depend on the number of samples used for training, and how they converge when the training set is larger than about 2500 samples.

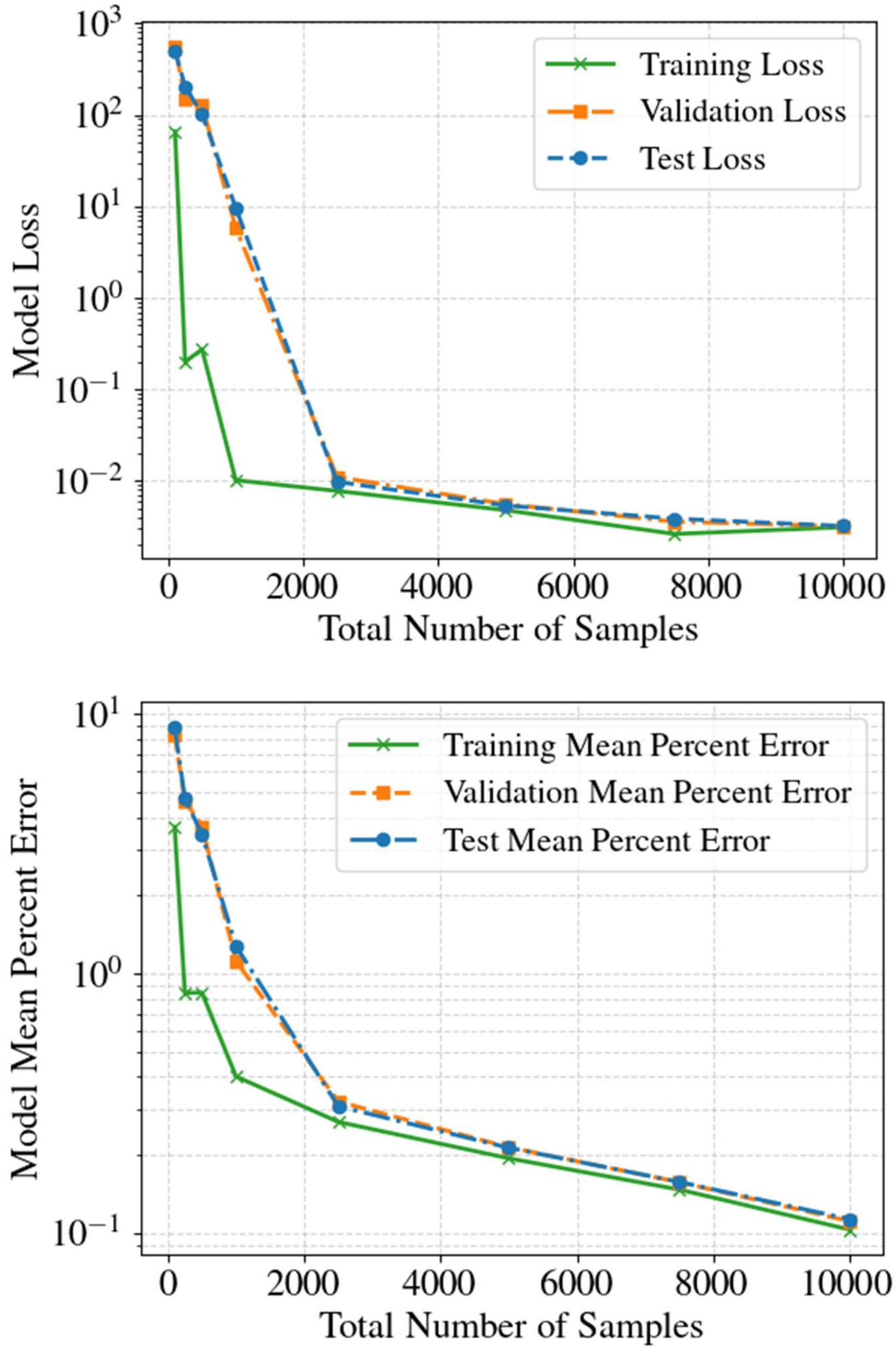

Figure 7. Training curves versus dataset size for the TNN. The two plots show the loss (top) and the mean percentage error (bottom) for the training, validation, and test sets vs number of samples used to train the model. The y-axes use logarithmic scaling, and the ratio of test to validation to training data was kept the same.

The final dataset included 10,000 samples, along with the 204 boundary configurations (described in Section II.A.2), split into training, validation, and test sets. The complete dataset had training, validation, and test losses of 0.00313, 0.00321 and 0.00323 respectively, with corresponding mean percentage error values of 0.103%, 0.111%, and 0.112%. These tightly clustered metrics indicate that the model generalizes well with minimal overfitting.

## IV. CONCLUSION

Three Artificial Neural Network (ANN) models were trained to predict self-shielded microscopic cross sections in the fuel region of a light water reactor pincell model, using only nuclide concentrations as inputs. The training data were generated using CASMO-8 group structure tallies derived from continuous-energy OpenMC pincell simulations. The fuel composition in each simulation featured a unique combination of 90 nuclides, with concentrations spanning enrichment bounds of 1.6% to 19.9% a/o and burnup levels from 18 to 210 MWd/kgIHM.

To improve accuracy, independent models were trained to predict quantities that a single general model consistently failed to capture accurately. A Fission Neural Network (FNN) predicted fission cross sections, a Total Neural Network (TNN) predicted the total, absorption, elastic scattering, and total scattering cross sections, and a targeted Plutonium Neural Network (PNN) was deployed specifically for resonances found in plutonium nuclides (group 4 for $^{239}$Pu and $^{241}$Pu; group 5 for $^{238}$Pu, $^{240}$Pu, and $^{242}$Pu). The combined models achieved an overall mean percentage error of 0.119%. An analysis of the residual distributions demonstrated that 93.0% of predictions fell within one standard deviation of the corresponding OpenMC Monte Carlo tally, and 97.6% of predictions were within a relative error of 1%. These results are consistent with findings from other groups [12, 13, 14,15, 17], who reported mean absolute percentage errors of less than 1% across various group structures and nuclide combinations. When the predicted microscopic cross sections were used to generate a macroscopic cross section library at runtime, the multigroup simulations achieved a mean absolute $\Delta k_{eff}$ of 200 pcm compared to the ground truth multigroup runs. This performance also compares well with other published works, which demonstrate average $\Delta k_{eff}$ between 398 and 965 pcm [15,17].

Across all test cases, only 0.0021% of the 5.4 million test case predictions exceeded a relative error of 10%. The most notable limitation was the prediction of the group 6 absorption cross section of $^{136}Xe$, which reached a maximum relative error of 119%. Targeted neural networks trained on specific groups and isotopes, similar to the PNN, could improve the predictive accuracy for $^{136}Xe$ and the other isotopes listed in Table 4.

The largest observed $\Delta k_{eff}$ between the ANN-predicted and OpenMC-generated cross section libraries was 1,189 pcm. Surprisingly, this test case did not have the largest cross section error. No clear correlation was found between cross section error (mean or mean absolute) and $\Delta k_{eff}$. For example, one test case with a mean absolute error of 0.4% the $\Delta k_{eff}$ was nearly 200 pcm, while a case with less than 0.1% error had a $\Delta k_{eff}$ over 700 pcm. Because $k_{eff}$ depends on the ratio of neutron production to loss, errors in the fission and absorption cross sections can either compound or cancel. For instance, if both fission and absorption cross sections are over predicted by similar percentages, their errors largely cancel and the ratio will remain relatively accurate. On the other hand, if one is underpredicted while the other is overpredicted, the errors compound and $\Delta k_{eff}$ increases substantially. We suspect that this phenomenon explains why cross section error and $\Delta k_{eff}$ appear to be uncorrelated. However, the error metrics examined here cannot explain the magnitude of the 1,189 pcm case. Further investigation is needed, such as a per-nuclide, per-reaction sensitivity analysis to identify which cross sections most strongly influence $\Delta k_{eff}$.

Currently, these models are limited in their applicability to a specific geometry and temperature profile. Future work will expand the models' scope to accommodate diverse reactor spectra, variable geometries, and temperatures. These can be implemented using additional input neurons encoding temperature, spectrum, and geometry, or by training additional neural networks that can refine a basic infinite medium calculation. Simple ways to encode geometry have a basis in well-established methods such as the Wigner Rational Approximation for convex geometries. The utility of the method also needs to be broadened by training neural network models to predict shielded quantities in a variety of commonly used group structures. Ultimately, a deployable method requires the ability to predict self-shielding in all nuclear data quantities for multigroup neutron transport, including scattering matrices.

**Disclosure Statement**

The authors report there are no competing interests to declare.

**Data Availability**

The data that support the findings of this study are openly available in zenodo at https://doi.org/10.5281/zenodo.21482578.

**Funding Details**

This work was supported by DOE-NE award DE-NE009418.

**ACKNOWLEDGEMENTS**

We thank Dr. Valerio Mascolino for his helpful suggestions.